\documentclass[submission, Phys]{SciPost}
\usepackage[english]{babel}
\usepackage{amsthm}
\usepackage{amsmath}
\usepackage{amssymb}
\usepackage{yhmath}
\usepackage[mathscr]{euscript}
\usepackage{titlesec}
\theoremstyle{definition}
\newtheorem{definition}{Definition}[section]

\theoremstyle{remark}

\usepackage{comment}

\begin{document}

\begin{center}{\Large \textbf{
Heat Kernel methods in the first-order formalism
}}\end{center}

\begin{center}
A. K. Mehta\textsuperscript{1}
\end{center}

\begin{center}
{\bf 1}Department of Physics, Kyunghee University, Seoul, Republic of Korea
\\
*abhishek.mehta@khu.ac.kr
\end{center}

\begin{center}
\today
\end{center}

\section*{Abstract}
{\bf In this paper, we extend the heat kernel methods to the first-order formalism of gravity, specifically, in the language of differential forms. This allows us to compute the effective dynamics of $4D$ gravity when the tetrad degrees of freedom are integrated out. We show that the resulting effective field theory is the Lorentz gauge theory. 


}

\vspace{10pt}
\noindent\rule{\textwidth}{1pt}
\tableofcontents\thispagestyle{fancy}
\noindent\rule{\textwidth}{1pt}
\vspace{10pt}
\section{Introduction}
The semi-classical approach to quantum gravity can be studied via two means - either by \emph{integrating out the metric} (MI)\cite{Freidel:2005me} or by \emph{integrating out the matter fields} (FI)\footnote{MI - Metric integration, FI - Field integration}
\begin{align}
    e^{-iS_{FI}[g]} = \int[\mathcal{D}\phi] e^{-i(S_{EH}[g]+S_{m}[g, \phi])}\label{EFTskt}\\ e^{-iS_{MI}[\phi]} = \int[\mathcal{D}g] e^{-i(S_{EH}[g]+S_{m}[g, \phi])}\label{SCAskt}
\end{align}
in the path integral. In the literature, the most popular approach is the FI method. For instance, if one computes the equation of motion of $S_{FI}[g]$, we obtain
\begin{align}
    &\int[\mathcal{D}\phi] \frac{1}{\sqrt{g}}\frac{\delta}{\delta g^{\mu\nu}}e^{-i(S_{EH}[g]+S_{m}[g, \phi])} = 0\notag\\&\implies  G_{\mu\nu}-8\pi\langle T_{\mu\nu}\rangle = 0\label{SCEE}
\end{align}
which is known as the \emph{semi-classical} Einstein's equation\cite{Ford:2005qz}, famously used to compute the Hawking radiation and black hole entropy\cite{dewitt1975quantum, Unruh:1974bw}. The FI approach is also used to compute corrections to an already known classical or quantum observable\cite{Schwartz:2014sze}. Notice that to solve Eq. \ref{SCEE}, the prescription that is used in the literature is to couple quantum fields to classical curved backgrounds. Therefore, another way to distinguish between these two approaches is by understanding that when we integrate out the matter fields, we are studying quantum fields coupled to curved classical backgrounds while when we integrate out the metric fields we are studying classical fields coupled to a quantum background. Since, half of the fields are classical in either of the appraoches, hence, the term \emph{semi-classical}. The MI approach as stated in Eq. \ref{SCAskt} is a largely unexplored territory in $4D$. But  computations in $3D$ quantum gravity\cite{Freidel:2005me} have shown that the MI effective action of $3D$ gravity is a noncommutative QFT which is a highly nontrivial result that cannot be captured by FI methods. Therefore, the MI approach as stated in Eq. \ref{SCAskt} may capture aspects of quantum gravity that are not obvious in FI approaches. This makes MI effective actions potentially a lot more exciting as it can shed light on some important nontrivial quantum gravity phenomena. Naturally, this approach is a lot more challenging as the path integral over the metric in $4D$ is mostly difficult. In $4D$, a full path integral over the metric can be achieved via LQG techniques\cite{rovelli2015covariant}. But it is not immediately obvious how such techniques may be utilized to derive the \emph{semi-classical approximations} of gravity via the MIs. The main reason for the popularity of the FI approach is due to the heat kernel method\cite{Decanini:2005gt}. The heat kernel method allows one to represent the functional determinants arising from the field path integration in terms of higher order curvature terms. That is how we know that any effective field theory of metric-gravity must necessarily involve higher order curvature terms. Therefore, in order to make the MI approach more computationally viable, we formulate an extension of the heat kernel formalism that will allow us to integrate out the metric fluctuations in a manner similar to the matter fields and allows us to derive the MI effective action to the Einstein-Cartan gravity.

\section{$4D$ gravity path integral}

The Einstein-Hilbert (EH) action in first-order formalism is given by
\begin{align}
    S  &= \frac{1}{2\pi G_N}\int_{\mathcal{M}}e^A \wedge e^B \wedge \widetilde{F}_{AB}~  
\end{align}
where
\begin{align}
    F^{AB} = d\omega^{AB}+\omega^A_{~B}\wedge\omega^{BC} \quad \widetilde{F}_{AB} = \epsilon_{ABCD}F^{CD}
\end{align}
where $\omega^{AB}$ is the connection 1-form and $e^A$ is the triad 1-form on the pseudo-Riemannian manifold $\mathcal{M}$. The equations of motions are obtained via the variation of the action with respect to 1-forms $e^{A}$ and $\omega^{AB}$ which are given by\footnote{The first reproduces the Einstein's field equations while the second is the torsionless constraint.}
\begin{align}
e^B \wedge \widetilde{F}_{AB} = 0 \quad De^A = 0
\end{align}
To compute our semi-classical approximation, we need to compute the following path integral
\begin{align}
Z[h, \omega] = \int [\mathcal{D}e] ~\exp\left({-\frac{i}{2\pi G_N}\int_{\mathcal{M}} \widetilde{F}_{AB}\wedge e^A \wedge e^B}-i\frac{\kappa}{2\pi}\int_{\mathcal{M}}e_A\wedge De^A\wedge *\mathcal{Q}_2\right)
\end{align}
where $\mathcal{Q}_2$ is the Chern-Simons 3-form given by
\begin{align}
    \mathcal{Q}_2 = \omega_A^{~~B} \wedge d\omega_B^{~~A} +\frac{2}{3}\omega_{A}^{~~B}\wedge\omega_{B}^{~~C}\wedge\omega_{C}^{~~A}
\end{align}
where the Hodge-$*$ operator is corresponding to an auxiliary metric $h_{\mu\nu}$ on the manifold. The Hodge operator works like a `Lagrange multiplier' that implements the torsionless constraint. The path integral measure $[\mathcal{D}e]$ is over the tetrad 1-form i.e. $e^A \equiv e^A_{\mu}dx^{\mu}$\cite{Carlip:1994tt}. The advantage of defining the path integral measure $[\mathcal{D}e]$ this way is that the measure automatically becomes invariant under diffeomorphism as the 1-form $e^A$ is naturally invariant under diffeomorphism. Since, the tetrad 1-form transforms like a vector field under local Lorentz transformations the path integral measure is also naturally invariant under local Lorentz transformations. However, the path integral measure $[\mathcal{D}e]$ is not invariant under local translations. This is not a problem because the action we are path integrating is not invariant under local translations. 
\begin{align}
&[\mathcal{D}e] \xrightarrow[\text{Local Lorentz}]{\text{Diffeomorphism}} [\mathcal{D}e]
\end{align}
Performing the path integral over $e^A$ to obtain
\begin{align}
&\int [\mathcal{D}e]\exp\left({-\frac{i}{2\pi G_N}\int_{\mathcal{M}} \widetilde{F}_{AB}\wedge e^A \wedge e^B}-i\frac{\kappa}{2\pi}\int_{\mathcal{M}}e_A\wedge De^A\wedge *\mathcal{Q}_2 \right) \notag\\&= \prod_{p \in \mathcal{M}}~{\det}^{1/2}\left(\frac{\widetilde{F}(p)}{2\pi G_N}+\frac{\kappa}{2\pi}*\mathcal{Q}_2(p)\wedge D_p\right) \equiv \prod_{p \in \mathcal{M}}~{\det}^{1/2}\mathcal{F}\label{PIA}
\end{align}
In the above, we have used higher dimensional generalization of the Volterra-type product integrals\footnote{See Appendix \ref{volterra}}. We have also made use of the ``fermionic" statistics of the 1-form\cite{Stanford:2017thb, KumarMehta:2024mpf} $e^A$ i.e $e^A \wedge e^B = -e^B\wedge e^A$. 
\subsection{Heat kernel method}
To make sense of the RHS of Eq. \ref{PIA}, we need to define a Green's function for the tetrad 1-form `field'. We define it as
\begin{align}
  &(\mathcal{F} \cdot G)_{A}^{C} =  \left(\frac{\widetilde{F}_{AB}}{2\pi G_N}+\eta_{AB}\frac{\kappa}{2\pi}*\mathcal{Q}_2\wedge D\right)\wedge G^{BC}_{x, x'} = -*U_A^C[\omega](x, x')\delta(x, x')\label{def}\\
  &U^A_B[\omega](x, x') = \mathcal{P}\exp\left(-\int^{x'}_x \omega^A_{~~B}\right)
\end{align}
where $\delta(x, x') \in \mathcal{D}'(U)$ where $\mathcal{D}'(U)$ is the space of distributions over $U \subseteq \mathcal{M}$\cite{vananalysis}. In the above, the gauge link is added following \cite{Leupold:2003zj}.
Before we proceed further, we make the following definition.
\begin{definition}[Green's function biform]
The Green's function $G_{x, x'} \in \Omega^{(1, 1)}(\mathcal{M}\times\mathcal{M})$ where $\Omega^{(1, 1)}(\mathcal{M}\times\mathcal{M})$ is the space of all biforms\cite{Ben_Sa_d_2019} with bidegree $(1, 1)$ on the product manifold $\mathcal{M}\times\mathcal{M}$.\label{propdef}
\end{definition}
We will explain why this definition is important in a later section\footnote{See Section \ref{tetprop}}. Symbolically, one may now write
\begin{align}
G^{BC}_{x, x'} &= -\left(\frac{\widetilde{F}_{AB}}{2\pi G_N}+\eta_{AB}\frac{\kappa}{2\pi}*\mathcal{Q}_2\wedge D\right)^{-1}*U_A^C[\omega](x, x')\delta(x, x')\equiv -\widehat{\mathcal{F}}^{-1}*U[\omega](x, x')\delta(x, x')\notag\\&= \left(\frac{\widetilde{F}_{AB}}{2\pi G_N}+\eta_{AB}\frac{\kappa}{2\pi}*\mathcal{Q}_2\wedge D\right)^{-1}*^{-1}U_A^C[\omega](x, x')\delta(x, x')
\notag\\
&=\left[\frac{*\widetilde{F}_{AB}}{2\pi G_N}+\eta_{AB}\frac{\kappa}{2\pi}*(*\mathcal{Q}_2\wedge D)\right]^{-1}U_A^C[\omega](x, x')\delta(x, x')\equiv \widehat{*\mathcal{F}}^{-1}U[\omega](x, x')\delta(x, x')
\end{align}
where we have used 
\begin{align}
    *^{-1} = -*
\end{align}
for a Lorentzian manifold in the above\cite{Nakahara:2003nw}. Now, we define the following heat kernel representation for the Green's function\footnote{From here onwards, the Latin indicies are implicit to avoid clutter.}
\begin{align}
   G \equiv *\mathcal{F}^{-1}=-\int_0^{\infty} d s K(s) \quad   K(s)\big[U[\omega]\delta(x, x')\big]\equiv \sinh[s \widehat{*\mathcal{F}}]~[U[\omega]\delta(x, x')\big] \equiv K(s|x, x')\label{HK}
\end{align}
where\footnote{We define the operation of as $*\widehat{\mathcal{F}}$ on any $f \in \Omega(\mathcal{M})$ as $*\widehat{\mathcal{F}}f = *(\widehat{\mathcal{F}}\wedge f)$} 
\begin{align}
    \sinh[s\widehat{*\mathcal{F}}] = \frac{ s}{2\pi G_N} *\widehat{\mathcal{F}} + \frac{1}{3!}\left(\frac{ s}{2\pi G_N}\right)^3 *(\widehat{\mathcal{F}}\wedge*(\widehat{\mathcal{F}}\wedge*\widehat{\mathcal{F}}))+ O(s^5)
\end{align}
We have made use of the following integral above
\begin{align}
    \int_0^{1/\epsilon}ds \sinh(a s) = -a^{-1} + \frac{1}{\epsilon} \quad \epsilon \to 0 \label{intrep}
\end{align}
neglecting the $1/\epsilon$ divergence as it is just a constant independent of $a$\cite{Vassilevich:2003xt}. Notice that the above power series when acting on the distribution $\delta(x, x')$ can be rewritten as an element of $\Omega^{(1, 1)}(\mathcal{M}\times\mathcal{M})$ so it is consistent with Definition \ref{propdef}. Now, consider the object
\begin{align}
\cosh[s\widehat{*\mathcal{F}}] = \eta_{AB} + \frac{1}{2!}\left(\frac{ s}{2\pi G_N}\right)^2 *(\widehat{\mathcal{F}}\wedge*\widehat{\mathcal{F}})_{AB} + O(s^4)
\end{align}
where each term in the above series is a 0-form when acting on the distribution $\delta(x, x')$ which is needed to define\footnote{The definition of $\operatorname{Tr}$ implicitly contains the Hodge-* operator.}
\begin{align}
   &\ln\operatorname{det}*\mathcal{F} = \operatorname{Tr}\ln*\mathcal{F} = -\int_0^{\infty} \frac{ds}{s} \operatorname{Tr}\cosh[s\widehat{*\mathcal{F}}] \equiv \int_0^{\infty} \frac{ds}{s} \operatorname{Tr}\mathcal{K}(s)\notag\\
   &\mathcal{K}(s|x, x') \equiv \mathcal{K}(s)[U[\omega]\delta(x, x')\big]\label{hkern}
\end{align}
In the above, we have made use of the following integral
\begin{align}
    \int_0^{1/\epsilon}\frac{ds}{s} \cosh(a s) = -\ln a + \frac{a}{\epsilon} \quad \epsilon \to 0 \label{intrep2}
\end{align}
which is simply an integral with respect to $a$ of Eq. \ref{intrep}. However, there is a problematic second term in the above which is dependent in $a$. But that will vanish due to the trace and the coincidence limit. To see this, look at
\begin{align}
    \widehat{\mathcal{F}} U[\omega](x, x')\delta(x, x') =  &\lim_{s\to 0}\left(\frac{\widetilde{F}_{AB}}{2\pi G_N}+\eta_{AB}\frac{\kappa}{2\pi}*\mathcal{Q}_2\wedge D\right) U[x, x'] \frac{1}{(4 \pi s^2)^{d / 2}} e^{-\sigma_h(x, x') / 2 s^2-s^2 m^2}\notag\\
    =  &\frac{\widetilde{F}_{AB}}{2\pi G_N}\delta(x, x')-\frac{\kappa}{4\pi}*\mathcal{Q}_2\wedge i_X F_{AB}\delta(x, x') \notag\\&-  U_{AB}[x, x']\kappa*\mathcal{Q}_2\wedge D\sigma_h \lim_{s\to 0} \frac{1}{(4 \pi s^2)^{d / 2 +1}} e^{-\sigma_h(x, x') / 2 s^2-s^2 m^2}
\end{align}
where $X^{\mu} = (x-x')^{\mu}$ and in the above we made use of\cite{Decanini:2005gt}
\begin{align}
    \delta(x, x') = \frac{1}{(4 \pi s^2)^{d / 2}} e^{-\sigma_h(x, x') / 2 s^2-s^2 m^2}
\end{align}
The first and second term vanishes due to the trace and the second term vanishes due $D\sigma_h \to 0$ in the coincidence limit\cite{Poisson:2011nh}. Hence, in Eq. \ref{intrep2}, the second term can be dropped as well. Now, to Eq. \ref{hkern} and Eq. \ref{HK}, one can associate the following heat equations
\begin{align}
    &\frac{\partial^2}{\partial s^2}\mathcal{K}(s|x, x') = *\widehat{\mathcal{F}}^2\mathcal{K}(s|x, x')\label{HK1}\\
    &\frac{\partial^2}{\partial s^2}K(s|x, x') = *\widehat{\mathcal{F}}^2 K(s|x, x')\label{HK2}
\end{align}
Notice that the propagator and the functional determinant follow two separate heat equations. This is because the heat equation is a second-order PDE and therefore, will have two independent solutions. Since, we are more interested in the semi-classical approximation, we will focus on solving Eq. \ref{HK1}. Since, $\mathcal{K}(0| x, x') = \delta(x, x')$, we, therefore, work with the following ansatz following\cite{Decanini:2005gt}
\begin{align}
  \hat{\mathcal{K}}(s \mid x, x')=\frac{1}{(4 \pi s^2)^{d / 2}} e^{-\sigma_h(x, x') / 2 s^2-s^2 m^2} \sum_{n=0}^{\infty} s^{2n} \hat{a}_n(x, x')\label{ansatz}
\end{align}
where $a_n(x, x')$ are 0-forms. For $d = 4$, we get the following recurrence relation near the coincidence limit $x \to x'$
\begin{align}
    4m^4 \hat{a}_{n-1}-2m^2(4n-7)\hat{a}_n+2(2n-3)(n-1)\hat{a}_{n+1} = *(\hat{\mathcal{F}}\wedge*(\hat{\mathcal{F}}\wedge \hat{a}_n))\label{RR}
\end{align}
We solve the recurrence relation starting with $\hat{a}_0(x, x') = U[\omega](x, x')$ to obtain the heat kernel upto $n=1$ to obtain
\begin{align}
    \hat{a}_0 & =U[\omega][x, x']\notag\\
    \hat{a}_1 &=  \frac{1}{6} *(\widehat{\mathcal{F}}\wedge *(\widehat{\mathcal{F}}\wedge \hat{a}_0)) -\frac{7}{3}m^2\hat{a}_0
    \label{f2a}
\end{align}
near the coincidence limit. However, notice that when $n=1$ the third term in the LHS of Eq. \ref{RR} vanishes making it impossible to determine $a_2$ from the recurrence relation, therefore, giving us a constraint relation close to the coincidence limit instead
\begin{align}
-10m^4\hat{a}_0 +\frac{10}{3}m^2*(\hat{\mathcal{F}}\wedge*(\hat{\mathcal{F}}\wedge \hat{a}_0)) = \frac{1}{6}*(\hat{\mathcal{F}}\wedge*(\hat{\mathcal{F}}\wedge *(\hat{\mathcal{F}}\wedge*(\hat{\mathcal{F}}\wedge \hat{a}_0))))\label{constrn}
\end{align}
However, using the above constraint we can inductively see that
\begin{align}
    \hat{a}_n = \mu_n \hat{a}_0 +\lambda_n *(\widehat{\mathcal{F}}\wedge*(\widehat{\mathcal{F}}\wedge \hat{a}_0))\label{ansatzR}
\end{align}
Using this ansatz in the recurrence relation in Eq. \ref{RR} with the constraint, we obtain a recurrence relation for $\mu_n$ and $\lambda_n$
\begin{align}
    &2m^4 \mu_{n-1}-m^2(4n-7)\mu_n+(2n-3)(n-1)\mu_{n+1} = -30m^4\lambda_n\notag\\
    &4m^4 \lambda_{n-1}-2m^2(4n+3)\lambda_n+2(2n-3)(n-1)\lambda_{n+1} = \mu_n
    \label{RR2}
\end{align}
Even though the recurrence relation could, we could determine the coefficients $\hat{a}_n(x, x')$ close to the coincidence limit up to some coefficients $\mu_n$ and $\lambda_n$.

\subsection{Semi-classical action}
We now have all the ingredients to compute the semi-classical approximation.
\begin{align}
     &\operatorname{Tr}\ln\mathcal{F} =  \left[\sum_{n=0}^{\infty}\mu_{n}I_{n}(m^2)\right]\int_{\mathcal{M}}*1 +\left[\sum_{n=0}^{\infty}\lambda_{n}I_{n}(m^2)\right]\frac{1}{(2\pi G_N)^2}\int_{\mathcal{M}}\widetilde{F}\wedge*\widetilde{F}\label{EA}
\end{align}
where
\begin{align}
    I_{n}(\alpha)\equiv\int_0^{\infty}s^{2n-5} e^{-\alpha s^2}
\end{align}
Notice that the effective action is precisely the Lorentz gauge theory\cite{Borzou:2014cea} in some auxilliary background $h_{\mu\nu}$ i.e. it is a theory where $\omega^{AB}$ is the gauge 1-form in the adjoint representation of the $SO^{+}(1, 3)$ group. Since, $h_{\mu\nu}$ is an auxilliary field that was added by hand, it should be integrated out using the equation of motion for $h_{\mu\nu}$ given by
\begin{align}
&F_{\mu}^{~~\alpha}F_{\nu\alpha} - \frac{1}{4}h_{\mu\nu}F_{\alpha\beta}F^{\alpha\beta} = (\pi G_N)^2\Lambda h_{\mu\nu}\\
&\Lambda = \frac{\sum_{n=0}^{\infty}\mu_{n}I_{n}(m^2)}{\sum_{n=0}^{\infty}\lambda_{n}I_{n}(m^2)}\label{cosmocon}
\end{align}
However, notice that when taken trace both sides we get $\Lambda = 0$ which contradicts Eq. \ref{cosmocon}. Therefore, that means the variation of the effective action in Eq. \ref{EA} with respect to $h_{\mu\nu}$ is not defined which is only possible if $h_{\mu\nu}$ is some fixed background. This means that what we essentially integrated out were just the quantum fluctuations in some fixed background. This shows that unlike in lower spacetime dimensions, in $4D$, we can never integrate out the metric completely. We will always have residual gravity through the means of a fixed background. We will learn how to determine $h_{\mu\nu}$ by an alternate means in the next section.\\

\subsection{Tetrad propagator}\label{tetprop}
We will now compute the tetrad propagator. Since, $K(0| x, x') = 0$, we, therefore, work with the following ansatz
\begin{align}
  \hat{K}(s \mid x, x')=e^{-s^2 m^2} \sum_{n=0}^{\infty} s^{2n+1} \hat{b}_n(x, x')\label{ansatz2}
\end{align}
Using Eq. \ref{ansatz2} in Eq. \ref{HK2}, one can derive the following recurrence relation close to the coincidence limit $x \to x'$
\begin{align}
    4m^4 \hat{b}_{n-1}-2m^2(4n+3)\hat{b}_n+2(n+1)(2n+3)\hat{b}_{n+1} = *(\hat{\mathcal{F}}\wedge*(\hat{\mathcal{F}}\wedge \hat{b}_n))\label{recurprop}
\end{align}
To solve the above recurrence, we require $\hat{b}_0$. To determine that, we make use of
\begin{align}
   \frac{\partial}{\partial s}K(s|x, x') = *\widehat{\mathcal{F}} \mathcal{K}(s|x, x')
\end{align}
Near the coincidence limit, we find that
\begin{align}
    -2m^2 \hat{b}_{n-3}+(2n-3)\hat{b}_{n-2} = \frac{*(\mathcal{\widehat{F}}\wedge\hat{a}_n)}{16\pi^2} \label{constrn3}
\end{align}
which at $n=2$ gives
\begin{align}
    \hat{b}_0 = \frac{*(\mathcal{\widehat{F}}\wedge\hat{a}_2)}{16\pi^2} = \frac{1}{16\pi^2}\left[\mu_2*(\mathcal{\widehat{F}}\wedge\hat{a}_0)+\lambda_2*(\mathcal{\widehat{F}}\wedge*(\mathcal{\widehat{F}}\wedge *(\mathcal{\widehat{F}}\wedge \hat{a}_0)))\right]
\end{align}
where we made use of Eq. \ref{ansatzR} in the above. Again inductively, we can surmise that
\begin{align}
    \hat{b}_n = \alpha_n *(\mathcal{\widehat{F}}\wedge\hat{a}_0) + \beta_n *(\mathcal{\widehat{F}}\wedge*(\mathcal{\widehat{F}}\wedge *(\mathcal{\widehat{F}}\wedge \hat{a}_0)))
\end{align}
Using the above in Eq. \ref{recurprop} with Eq. \ref{constrn}, we obtain
\begin{align}
    &4m^4\alpha_{n-1}-2m^2(4n+3)\alpha_n+2(n+1)(2n+3)\alpha_{n+1} = -60m^4\beta_n\notag\\
    &4m^4\beta_{n-1}-2m^2(4n+3)\beta_n+2(n+1)(2n+3)\beta_{n+1} = \alpha_n+20m^2\beta_n
\end{align}
Using the above, close to the coincidence limit, we can write the 2-pt function for the tetrad 1-form `fields' as 
\begin{align}
    \langle e^A(x) e^B(x')\rangle = G^{AB}_{x, x'} \underset{x \to x'}{=} \left[\sum_{n=0}^{\infty}\alpha_n I_{n+3}(m^2)\right] *(\mathcal{\widehat{F}}\wedge\hat{a}_0) + \left[\sum_{n=0}^{\infty}\beta_n I_{n+3}(m^2)\right]*(\mathcal{\widehat{F}}\wedge*(\mathcal{\widehat{F}}\wedge *(\mathcal{\widehat{F}}\wedge \hat{a}_0)))
\end{align}
In the coincidence limit, one can show the following
\begin{align}
    \langle e^{(A}(x) \wedge e^{B)}(x)\rangle = &\frac{C(m^2)\kappa}{8\pi^3G^2_N}\left[*(\widetilde{F}\wedge *(*\mathcal{Q}_{2}\wedge D*\widetilde{F})) + *\mathcal{Q}_{2}\wedge D*(\widetilde{F}\wedge *\widetilde{F})  \right]^{(AB)}\notag\\&+\frac{C(m^2)\kappa^2}{8\pi^4G^2_N}\left[ *\mathcal{Q}_2\wedge *(*\widetilde{F}\wedge *(*\mathcal{Q}_2\wedge *\widetilde{F}))\right]^{(AB)} \neq  0\quad C(m^2) = \left[\sum_{n=0}^{\infty}\beta_n I_{n+3}(m^2)\right]\label{2ptf}
\end{align}
where in the above we made use of\cite{Elze:1986qd}
\begin{align}
   &D_\mu(b) U[b, a]=-(b-a)^\nu \int_0^1 \mathrm{~d} s s U[b, z(s)] F_{\mu \nu}(z(s)) U[z(s), a]\notag\\
   &\implies D^2(b)U[b, a]\bigg|_{b\to a}
 = -\frac{F }{2}
 \end{align}
where $z(s) = a + (b-a)s$. Notice that this is a deviation from the expected
\begin{align}
    e^{(A}\wedge e^{B)}\equiv \frac{1}{2}(e^{A}\wedge e^{B} + e^{B}\wedge e^{A}) = 0
\end{align}
And the reason for this is that we are dealing with quantum geometries where the usual notion of differential geometry may not hold. This is to be expected as we have performed a path integral quantization of the tetrad 1-form and quantized differential forms have been shown to deviate from the traditional understanding of differential geometry\cite{KumarMehta:2024mpf}. However, any classical field will traverse such a quantum geometry as if it is a classical geometry. Therefore, in the ``classical geometry" limit of 
\begin{align}
     \langle e^{(A}(x) \wedge e^{B)}(x)\rangle\xrightarrow[\text{Geometry}]{\text{Classical}} 0
\end{align}
 Eq. \ref{2ptf} now becomes a constraint that along with the equation of motion for $\omega^{AB}$ in the action in Eq. \ref{EA} will determine $h_{\mu\nu}$. Notice that $h_{\mu\nu}$ is not determined by the Einstein's field equation but a nontrivial constraint arising from quantum gravity. Therefore, $h_{\mu\nu}$ represents the \emph{effective} classical geometry that is experienced by the classical fields, namely $\omega^{AB}$, in quantum backgrounds. This also explains why we failed to determine $h_{\mu\nu}$ from the variation of Eq. \ref{EA} in the previous section. Since, we are working with the MI approach, we are studying classical fields coupled to a quantum gravitational background so, naturally, it will not be determined by the equations of motions. 
\\\\
Now, we will motivate Definition \ref{propdef} and show how it ensures consistency of the formalism. Consider the following path integral
\begin{align}
    Z[J] = \int [\mathcal{D}e] \exp\left(-\frac{i}{2\pi G_N}\int_{\mathcal{M}}\mathcal{F}_{AB}\wedge e^A\wedge e^B-i\int_{\mathcal{M}} e^A\wedge J_B\right)
\end{align}
where $J_B$ is a source 3-form. We can can rewrite the above as
\small
\begin{align}
    Z[J] = \int [\mathcal{D}e] \exp\left(-\frac{i}{2\pi G_N}\int_{\times^3\mathcal{M}}[*\delta(x) e^A(y) + \pi G_N J_C(x)\wedge G^{CA}_{x,y}]\wedge \mathcal{F}_{AB}(y)\wedge [e^B(y) *\delta(x')+\pi G_N G^{BD}_{y,x'}\wedge J_D(x')]\right.\notag\\\left.-\frac{i\pi G_N}{2}\int_{\times^2\mathcal{M}} J_A(x)\wedge G^{AB}_{x,x'}\wedge J_B(x')\right)
\end{align}
\normalsize
where $\times^n\mathcal{M}$ is a product manifold of $\mathcal{M}$ with itself $n$-times. One can redefine the tetrad 1-form 'field' to\footnote{$\mathcal{M}_x$ represents the integral over the $x$-coordinates.} 
\begin{align}
    e^A(y) \to  \int_{\mathcal{M}_x}[*\delta(x) e^A(y) + \pi G_N J_C(x)\wedge G^{CA}_{x,y}] \equiv e^A(y) + \pi G_N\phi^A(y) 
\end{align}
where $\phi^A$ is some 1-form on $\mathcal{M}$ and compute the above gaussian integral to obtain
\begin{align}
    Z[J] = \exp\left[-\frac{i\pi G_N}{2}\int_{\mathcal{M}\times\mathcal{M}} J_A(x)\wedge G^{AB}_{x,x'}\wedge J_B(x')\right]\prod_{p\in\mathcal{M}}{\det}^{1/2}\left[\frac{\mathcal{F}(p)}{2\pi G_N}\right]  
\end{align}
Notice that the above makes sense only when $G^{AB}_{x,x'}$ is a $(1,1)$-bidifferential form on the manifold $\mathcal{M}\times\mathcal{M}$. If $G^{AB}_{x,x'}$ were just an element of $\Omega^2(\mathcal{M})$, then the last term in the exponential would have vanished.  Therefore, Definition \ref{propdef} ensures that the above path integral computation holds and  implying that the tetrad propagator is a well-defined observable.
\newpage

\section{Summary and Discussion}
In this paper, we first defined a diffeomorphic, local Lorentz invariant path integral measure over the tetrad 1-form $e^A$ which reduced the path integral to a simple ``fermionic" gaussian integral and allowed us to compute the functional determinant of the operator $\mathcal{F}$ which is essentially a derivative operator such that $\mathcal{F}:\Omega^r(\mathcal{M})\to\Omega^{r+2}(\mathcal{M}) $. To make sense of the functional determinant of such an operator, we represented it as the integral of the heat kernel $\mathcal{K}(s)$. The trace of the $\mathcal{K}(s)$ is naturally a 4-form which is necessary for it to produce the  semi-classical approximate action. We solve the corresponding heat equation for the kernel matrix $\mathcal{K}(s|x, x')$ and determine that the semi-classical effective action is precisely the Lorentz gauge theory with a cosmological constant in some fixed metric background $h_{\mu\nu}$. We also computed the 2-pt function of the tetrad form close to and at the coincidence limit and using the symmetrized 2-pt function demonstrated how the wedge product doesn't anticommute like it is supposed to signifying a quantization of the geometry. We later defined the ``classical geometry" limit as the vanishing of the symmetrized 2-pt function at the coincidence limit which lead to a constraint that along with the equation of motion for the gauge field of the Lorentz gauge theory can be used to determine $h_{\mu\nu}$. \\

The obvious follow-up direction one can undertake after this is computing the observables beyond the coincidence limit. That will make this formalism mathematically more complete. Another obvious follow-up would be to solve for $h_{\mu\nu}$ and study its properties. This will teach us a lot about the effect of quantum geometries on classical fields and perhaps, may give some insight on experimental detection of quantum geometries. Notice how in this exercise we always worked with differential forms because this exercise, apart from an attempt at deriving the semi-classical approximation to the Einstein-Cartan gravity, was also an attempt at doing physics at the level of the mathematical formulation itself. Usually, in the study of gravity, the gravitational action is written in the component form before implementing the field theoretic techniques. This adds a lot of computational challenges to a problem that is conceptually hard to begin with. Hence, in the study of quantum gravity, it would be very convenient if we can develop field theoretic tools and techniques that can allow us to do physics at the level of differential forms where the general coordinate invariance is manifest. In fact, in the literature one can find many instances where computations in quantum theories of gravity are done with this spirit\cite{Carlip:1994tt, Stanford:2017thb, Witten:1988hf, KumarMehta:2024mpf} without explicit acknowledgment for the need of such formulations. Also, notice that in our study we did not couple any matter to gravity. The reason for that is any matter that is coupled to gravity necessarily involves the Hodge-* operator with respect to the dynamical background metric field of the theory and there is no well-known procedure to integrate them out if they appear in the action. This is perhaps most important problem in this approach to quantum gravity is a part of our ongoing investigation and we will certainly report any development in this regard in the future.
\section*{Acknowledgements}
I would like to acknowledge the moral support of Hare Krishna Movement, Pune, India. This research is dedicated to the people of India and the Republic of Korea for their steady support of research in theoretical science. This work was supported by the NRF grant funded by the Korea government (MSIT) (No. 2022R1A2C1003182).
\appendix
\section{Invariance under local translations}
Consider the following variation of the tetrad 1-form and the connection 1-form under local translations\cite{Carlip:1994tt}
\begin{align}
&\delta e^{a} =D\tau^a \\
&\delta \omega^{ab} =0
\end{align} 
Hence, we have 
\begin{align}
\delta S = 2\int_{\mathcal{M}} \star F_{ab}\wedge e^a \wedge \delta e^b = 2\int_{\mathcal{M}} \star F_{ab}\wedge e^a \wedge D\tau^b
\end{align}
Now, we use the identity
\begin{align}
0 = \int_{\mathcal{M}}\star d(F_{ab}\wedge e^a\wedge \tau^b) =  \int_{\mathcal{M}}\star [F_{ab}\wedge De^a\wedge \tau^b+F_{ab}\wedge e^a\wedge D\tau^b]
\end{align}
to rewrite the variation as follows
\begin{align}
\delta S = -2\int_{\mathcal{M}} \star F_{ab}\wedge De^a \wedge \tau^b
\end{align}
We immediately notice that for invariance under local translations, we should have        $De^a = 0$, which is precisely the torsionless condition. However, we are working with the possibility of spacetime with torsion, so we do not impose the torsionless condition.

\section{Higher-dimensional Volterra-type product integral}\label{volterra}
Consider the product integral orignally proposed by Volterra\cite{volterra1938operations}
\begin{align}
\prod_a^b (1+f(x)dx) = \exp\left(\int_a^b f(x) dx\right)
\end{align}
Now, interpreting $f(x)dx$ as the top-form on a $1D$-manifold with the integral open set $(a, b)$, we may say that
\begin{align}
e^{f(x)dx} = 1+f(x)dx
\end{align}
which we make use of in the LHS and rewrite the product integral as
\begin{align}
\prod_a^b (1+f(x)dx) = \prod_a^b e^{f(x)dx} =\exp\left(\int_a^b f(x) dx\right)
\end{align}
Now, we may similarly generalize the product integral to top-forms $\Omega$ on $d$-dimensional manifolds $\mathcal{M}$ in an open set $U$ as
\begin{align}
\prod_{p\in U} (1+\Omega(p)) = \prod_{p \in U} e^{\Omega(p)} =\exp\left(\int_{U \subset \mathcal{M}} \Omega\right)
\end{align} 
\bibliography{QGBib}
\bibliographystyle{unsrt}
\end{document}